\documentclass[fleqn,usenatbib]{mnras}
\usepackage{newtxtext,newtxmath}
\usepackage[T1]{fontenc}
\DeclareRobustCommand{\VAN}[3]{#2}
\let\VANthebibliography\thebibliography
\def\thebibliography{\DeclareRobustCommand{\VAN}[3]{##3}\VANthebibliography}
\usepackage{graphicx}	% Including figure files
\usepackage{amsmath}	% Advanced maths commands

\usepackage{amssymb}	% Extra maths symbols
\usepackage{gensymb}	% use \degree
\usepackage{threeparttable}  % table notes
\usepackage{ulem} % use delete symbol \sout{}
\title[The rotation curve of M31]{The rotation curve and mass distribution of M31}

\author[Zhang et al.]{
Xiangwei Zhang,$^{1}$\thanks{E-mail: zhangxw@mail.ynu.edu.cn}
Bingqiu Chen,$^{2}$\thanks{E-mail: bchen@ynu.edu.cn}
Pinjian Chen,$^{1}$
Jiarui Sun,$^{1}$
and Zhijia Tian$^{1}$
\\
$^{1}$Department of Astronomy, Yunnan University, Kunming, 650500, P. R. China\\
$^{2}$South-Western Institute for Astronomy Research, Yunnan University, Kunming, 650500, P. R. China
}

\date{Accepted XXX. Received YYY; in original form ZZZ}

\pubyear{2023}

\begin{document}
\label{firstpage}
\pagerange{\pageref{firstpage}--\pageref{lastpage}}
\maketitle

\begin{abstract}
To gain a better understanding of the Andromeda galaxy M31 and its role in the Local Group, measuring its mass precisely is essential. In this work, we have constructed the rotation curve of M31 out to $\sim$125\,kpc using 13,679 M31 objects obtained from various sources, including the LAMOST data release 9 (LAMOST DR9), the DESI survey, and relevant literature. We divide all objects in our sample into bulge, disk and halo components. For the sources in the M31 disk, we have measured their circular velocities by a kinematic model with asymmetric drift corrections. For the bulge and halo objects, we calculate their velocity dispersions and use the spherical and projected Jeans equation to obtain the circular velocities. Our findings indicate a nearly isotropic nature for the M31 bulge, while the halo exhibits tangential anisotropy. The results show that the rotation curve remains constant at $\sim$220\,km\,s$^{-1}$ up to radius $\sim$25\,kpc and gradually decreases to $\sim$170\,km\,s$^{-1}$ further out. Based on the newly determined rotation curve, we have constructed a mass distribution model for M31. Our measurement of the M31 virial mass is $M_{\rm vir} = 1.14^{+0.51}_{-0.35} \times 10^{12} M_\odot$ within $r_{\rm vir} = 220 \pm 25$\,kpc. 
\end{abstract}


\section{Introduction}

Out of the billions of galaxies in the observable universe, only a few galaxies in the Local Group can be resolved into individual objects and studied in full phase space such as the positions, velocities and element abundances. The Andromeda galaxy M31, being the nearest spiral dynamical system to us and the only large spiral galaxy that can be studied in full and detail, serves as an ideal astrophysical laboratory for exploring the formation and evolution of galaxies.

Mass is a fundamental property of a galaxy. An accurate measurement of the mass of M31 is crucial for understanding its matter distribution,  formation and evolution history, and the role played in the Local Group. So far, there are a number of efforts devoted to the determination of the dynamical mass of M31, with a variety of tracers such as emission-line objects \citep[e.g.][]{Rubin1970,Kafle2018}, globular clusters \citep[e.g.][]{Galleti2006,Lee2008,Veljanoski2014}, tidal stellar streams \citep[e.g.][]{Ibata2004,Fardal2013,Dey2022} and satellite galaxies \citep[e.g.][]{cote2000,Watkins2010,Tollerud2012}; and a variety of methods such as the rotation curve method \citep[e.g.][]{Rubin1970,Carignan2006,Chemin2009,Sofue2015}, velocity distribution fitting \citep[e.g.][]{Evans2000,Watkins2013}, virial theorem \citep[e.g.][]{Hartwick1974,Tollerud2012}, projected mass estimator and tracer mass estimator \citep[e.g.][]{Federici1990,Federici1993,Lee2008,Veljanoski2013}. 

\begin{figure}
	\includegraphics[width=\columnwidth]{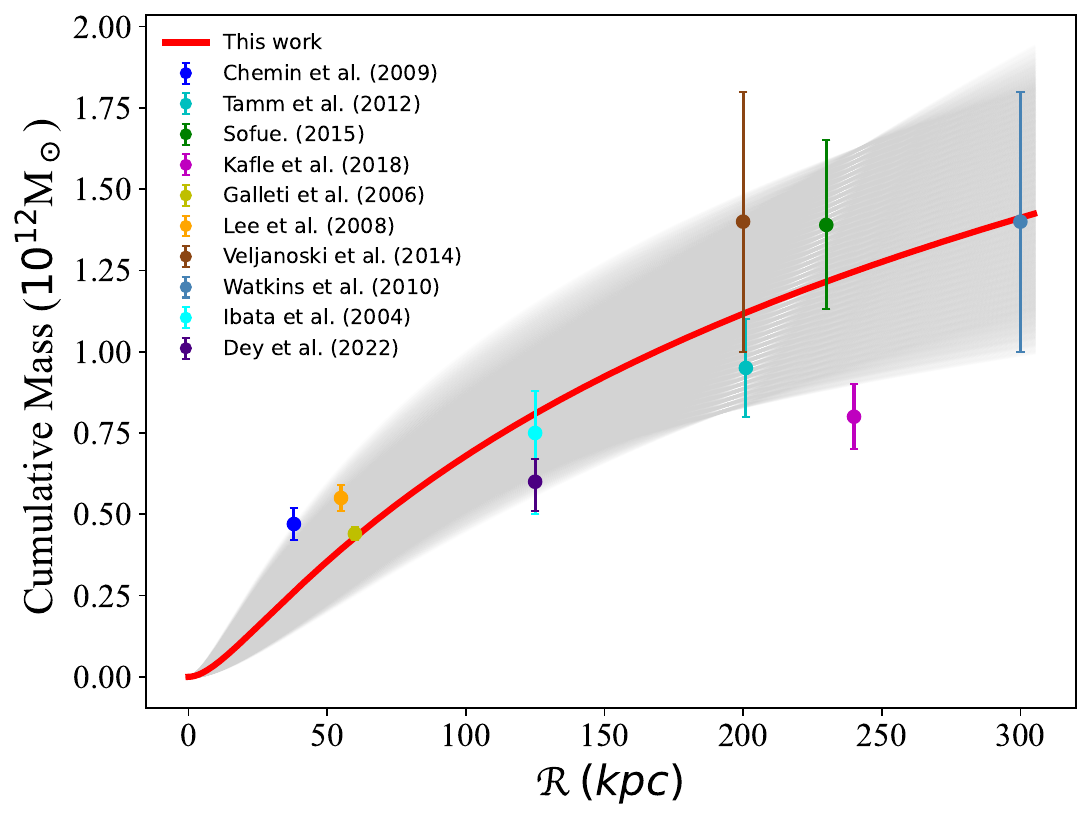}
	\caption{A summary of recent M31 mass measurements from previous works. The cumulative mass profile calculated in this study is represented by the red solid line with grey shade indicating uncertainties.}
	\label{fig:CumulativeMass}
\end{figure}

Even with these extensive efforts to determine the mass of M31, an accurate estimation still remains elusive. A summary of previous mass measurements is presented in Fig.~\ref{fig:CumulativeMass}, indicating significant uncertainties in many of the results due to unclear parameter values like the halo anisotropy parameter and limited data sample sizes. Intriguingly, it is still uncertain whether M31 or the Milky Way is the most massive member of the Local Group \citep[e.g.][]{Evans2000,Watkins2010}. 

Most of the mass measurements of M31 are based on the line-of-sight (LOS) velocity measurements of objects in it. Many projects have been undertaken to observe the spectra of objects in M31. \citet[hereafter M06]{Merrett2006} published a catalogue of 3,300 emission-line objects with their positions, magnitudes, and velocities, which were discovered using the Planetary Nebula Spectrograph. \citet{Halliday2006} presented a catalogue of 723 planetary nebulae (PNe), utilizing the WYFFOS fiber spectrograph. \citet{Galleti2009} presented LOS velocities of 118 M31 globular clusters from a series of spectroscopic observations \citep{Galleti2006,Galleti2007,Galleti2009}. 
\citet{Caldwell2009} used the Multi-Mirror Telescope (MMT) Hectospec to observe spectra of 670 cluster candidates. There are also other catalogues of emission-line objects \citep[e.g.][]{Sanders2012, Martin2018} and clusters \citep[e.g.][]{Veljanoski2013,Fan2011, Fan2012} in M31 that have been established. The Spectroscopic and Photometric Landscape of Andromeda’s Stellar Halo (SPLASH; \citealt{Guhathakurta2006}) survey utilized the DEIMOS instrument on the Keck II Telescope to study the kinematic and dynamical properties of the disk and halo of M31. More than 5,000 red giants in 50 fields have been observed so far \citep{Gilbert2018}. Most recently, the Dark Energy Spectroscopic Instrument (DESI) survey has provided LOS velocities of 7,617 M31 sources \citep{Dey2022}.

With 4000 fibers and a field of view of 20\,deg$^2$, the Large Sky Area Multi-Object Fiber Spectroscopic Telescope (LAMOST; \citealt{Cui2010,Cui2012}) serves as a powerful tool to observe object in M31. The LAMOST Spectroscopic Survey of the Galactic Anti-center (LSS-GAC; \citealt{Liu2014,Yuan2015}) systematically observed sources in M31, such as the PNe \citep[e.g.][]{Yuan2010,Xiang2017,Zhang2020}, H~{\sc ii} regions \citep[e.g.][]{Zhang2020,Alexeeva2022}, supergiants \citep[e.g.][]{Huang2019,Liu2022} and globular clusters \citep[e.g.][]{Chen2015,Chen2016,Wang2021}.
% use of latex cite: https://blog.csdn.net/jueshu/article/details/89186918

We have attempted to compile the most comprehensive catalogue of sources in M31 with LOS velocity measurements by utilizing data from these projects. This promising sample will enable us to accurately determine the mass of M31. This paper is the first in a series of articles in which we will establish the rotation curve of M31. The rotation curve is a crucial tool for constraining the mass distribution of the galaxy, including its dark matter content \citep[e.g.][]{Rubin1970,Carignan2006,Chemin2009}.

\section{Data}
\label{sec:Data}

We hereby introduce the data set used in the current work. The sample comprises objects primarily selected from two sources, M31 objects identified from the LAMOST data and existing ones from the literature.

\subsection{Sources selected from the LAMOST data}
\label{sec:M31 Membership}

In this work, we select objects in M31 from the LAMOST DR9 low resolution catalogue \citep[][]{Luo2015} released in March 2022. Based on 10 years of spectroscopic surveys, the LAMOST DR9 low-resolution catalogue contains 11.22 million spectra with a resolution of $R\sim1800$ and a wavelength range of 3800 -- 9000\,\AA. We briefly describe here how we selected the sources in M31 from LAMOST. For more details we refer the reader to Chen et al. (in prep.). 

To identify objects in M31, we initially search for all sources within a sky area of $\pm$15\degr centered on M31 from the LAMOST DR9. This yields a sample of 648,485 spectra. To distinguish genuine M31 members, we first use a machine learning algorithm to classify these objects. We cross match the LAMOST sources with the SIMBAD database and catalogues from previous studies. We adopt objects with confirmed types as our training sample. Using the fluxes of the individual wavelengths as the input parameters, we train a random forest classifier to categorize the LAMOST objects into three classes: objects with emission lines, stars without emission lines, and star clusters. Spectra with significant emission lines and no obvious continuum are selected as candidates for emission-line nebulae. The nebulae candidates are then classified into various types such as PNe, H~{\sc ii} regions, and others based on their emission line flux ratios. This classification is carried out using the Baldwin-Phillips-Terlevich (BPT) diagram \citep[][]{Baldwin1981}. Spectra with significant emission lines and continuum are classified as stars with emission lines. Using a straightforward kinematic analysis similar to that of \citet{Drout2009}, we select M31 stars based on their positions and LOS velocities, both those with emission lines and those without. For star cluster candidates, we examine their morphology by analyzing their Pan-Andromeda Archaeological Survey \citep[PAndAS;][]{McConnachie2009} $g$- and $i$-band images. 

As a result, the LAMOST M31 source catalogue includes a total of 1,019 sources, consisting of 258 emission-line nebulae, 382 clusters and candidates, and 379 supergiants and candidates.

\subsection{Sources from the literature}
\label{sec:Extra M31 Samples}

The quantity of tracers has a greater impact on the accuracy and precision of mass measurement than the quality \citep[][]{Hughes2021}. In recent years, the number of spectroscopic observed sources in M31 has significantly increased, thanks in large part to the development of wide-field multi-object spectrographs. To expand the sample size, we have also incorporated previous M31 source catalogues that have LOS velocity measurements in our work. The sources in these catalogues are primarily stars \citep[referred to as D09, C10 and D23]{Drout2009,Cordiner2011,Dey2022}, emission-line objects \citep[referred to as M06, H06, S12, and M18]{Merrett2006,Halliday2006,Sanders2012,Martin2018}, and clusters  
\citep[referred to as C16, S16 and V14]{Chen2016,Sakari2016,Veljanoski2014}. Revised Bologna Catalogue Version 5 \citep[RBCV5;][]{Galleti2004}\footnote{http://www.bo.astro.it/M31/} and Studies of Resolved Objects in M31\footnote{https://lweb.cfa.harvard.edu/oir/eg/m31clusters/M31\_Hectospec.html} by \citet[]{Caldwell2016} are also included (referred to as RBCV5 and Caldwell). We have compiled all published catalogues from the literature that we are aware of. The information pertaining to all collected catalogues from previous studies is presented in Table~\ref{tab: extra data information}, including the facilities used in observation, fiber sizes, the wavelength coverage and resolution of the spectra, the types and size of samples.

\begin{table*}
	\centering
	\caption{Information of our collected literature catalogues.}
	\label{tab: extra data information}
	\resizebox{\linewidth}{!}{  % Here "!" means adaptive scaling based on the aspect ratio
	\begin{tabular}{lcccccr} 
		\hline
		Catalogue & Spectrograph/Telescope& Fiber Size (arcsec) & Wavelength (\AA) & Resolution & zero point (km\,s$^{-1}$) &Note \\
		\hline
		M06  & PN.S/WHT      & -        & $\sim$5002.2 &-   &0.83 & 3,300 emission-line objects\\
		H06 & WYFFOS/WHT     & 2.7, 1.6& $\sim$4808 -- 5158 &$R\sim$5000, 10000 & 3.8& 723 PNe\\
		S12  & Hectospec/MMT & 1.5   & 3650 -- 9200  &$R\sim$1200 &$-$3.0 & 253 H~{\sc ii} regions and 407 PNe\\  
		%https://iopscience.iop.org/article/10.1086/497385
		M18   & SITELLE/CFHT  & -  & 6470 -- 6850 & $R$$\sim$5000 & $-$29.6&nearly 800 emission-line objects\\ 
		D09   & Hectospec/MMT & 1.5    & 4550 -- 7050 &$R\sim$2000   & $-$9.1& 116 supergiants\\
		C10  & GMOS/Gemini N & 1$^\text{a}$ &  $\sim$4000 -- 7000 & $R$$\sim$2500 -- 3500& -&34 OB stars\\
		C16     & LAMOST        & 3   & $\sim$4000 -- 9000 & $R\sim$1800 &-& 306 massive clusters\\
		S16   & APOGEE/APO    & 2   & $\sim$3000 -- 9000 & $R$$\sim$22500 &$-$0.35& 25 GCs\\ % https://ui.adsabs.harvard.edu/abs/2013AJ....146...81Z/abstract
		RBCV5                & - & - & - & -     & $-$8.29 & 956 clusters and candidates\\
		V14 & ISIS/WHT & 1.5-2$^\text{a}$ & $\sim$3500 -- 5100, $\sim$7500 -- 9200 & $R\sim$1500, 2700&$-$11.4& 78 clusters\\
		~ & RC/KPNO & 2$^\text{a}$ & $\sim$3500 -- 6500 &$R\sim$1300 &~& ~\\
		~ & GMOS/Gemini N& 0.75$^\text{a}$ & $\sim$7450 -- 9500 &$R\sim$4000 &~ &~\\
		D23     & DESI/Mayall telescope & 1.5 &3600 -- 9800&  $R\sim$2000 & $-$9.8&7,617 M31 sources\\ %7438+43+136
		Caldwell &Hectospec/MMT &1.5 &3650 -- 9200  &$R\sim$1200  &- & $\sim$5,000 emission-line objects, stars and clusters\\
		\hline
	\end{tabular}
	}
	\begin{tablenotes}
		\footnotesize
		\item $^\text{a}$Slit width
	\end{tablenotes}
\end{table*}

To combine all these catalogues, both position error and LOS velocity error should be considered. We conduct cross-matching between catalogues, selecting a radius based on the tracer types and fiber sizes. We adopt a matching radius of 1.5\,arcsec for stars and emission-line objects in M06, H06, S12, M18, D09, C10, D23, and Caldwell; whereas for clusters in C16, S16, V14, RBCV5, and Caldwell, we use a radius of 3\,arcsec. If a source appears in multiple catalogues, we prioritize its LOS velocity measurement and type with smaller velocity uncertainty or better observing condition. It is possible that there are systematic differences between velocities from different catalogues. We perform cross-matching between literature catalogues and the LAMOST M31 source catalogue. We obtain zero-point values from the velocities of the common sources, which are then utilized to correct the velocities of all sources in the literature catalogues and align them with those from LAMOST. In Table~\ref{tab: extra data information}, we additionally present the computed zero-point values for the individual catalogues.

\begin{figure*}
	\includegraphics[width=1.4\columnwidth]{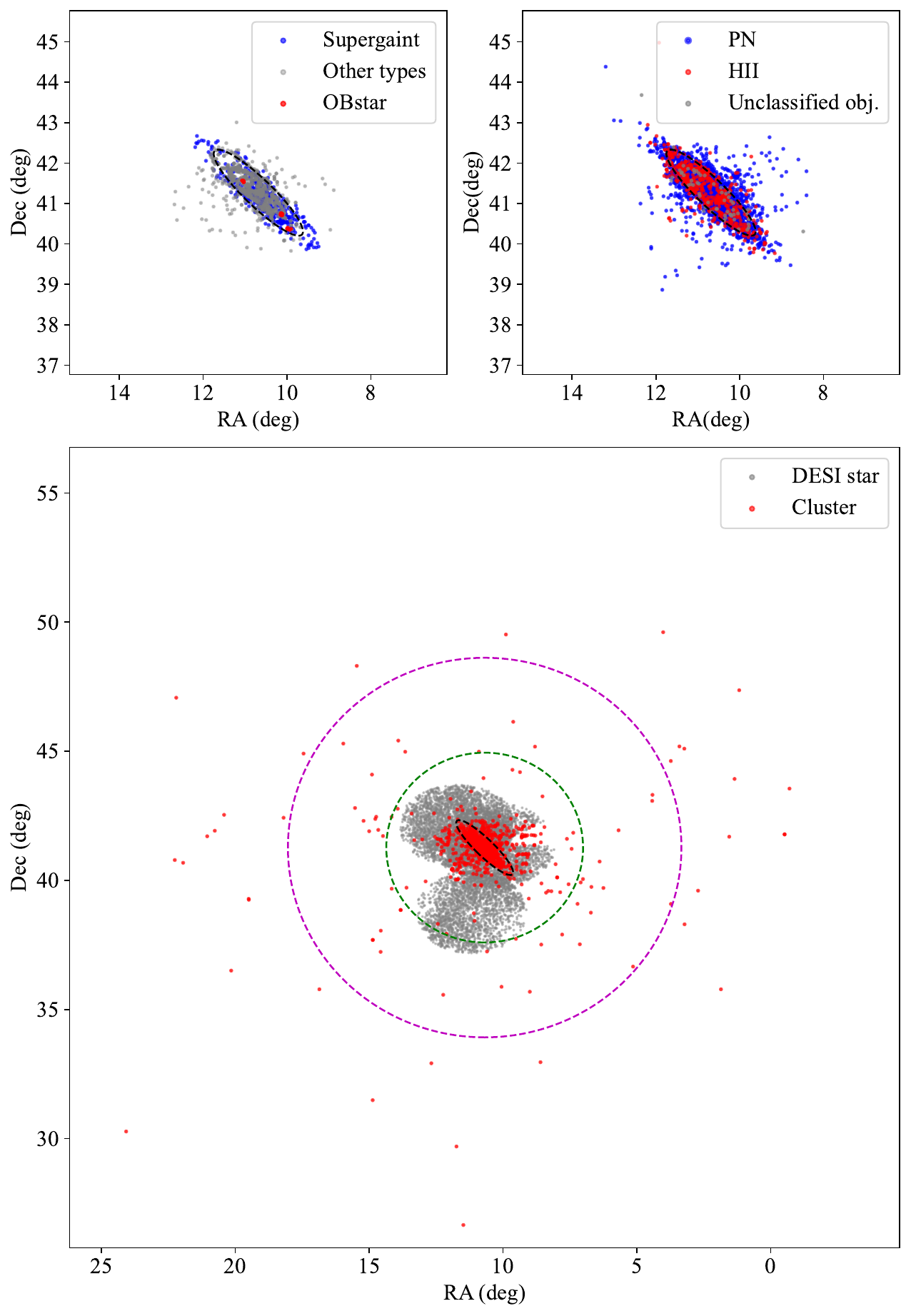}
    \caption{The spatial distribution of all sources in our final catalogue. The upper left panel illustrates the distribution of stars in the disk of M31, which includes OB stars (red dots), supergiants (blue dots), and other types (gray dots). The upper right panel depicts the distribution of emission-line objects, including PNe (blue dots), H~{\sc ii} regions (red dots), and unclassified objects (gray dots). The bottom panel shows the distribution of clusters (red dots) and stars in the halo of M31 (gray dots). The black dashed ellipses represent circles with a disk projected radius $R = 20$\,kpc, while the green and purple dashed circles represent sky projected radii of $r_p$ = 50\,kpc and 100\,kpc, respectively.}
    \label{fig:SourceSpaceDistribution}
\end{figure*}

The resulting combined catalogue comprises 13,679 M31 sources, all of which have corresponding LOS velocity measurements. Among these sources, 8,341 are stars, with the majority originating from D23. There are 3,997 emission-line objects, which consist of 3,274 PNe and candidates, 661 H~{\sc ii} regions and candidates, and 62 sources with types listed as `other type' or `unknown'. Additionally, there are 1,103 clusters, the majority of which (902) are globular clusters located in the halo of M31. The spatial distribution of all the catalogued sources is shown in Fig.~\ref{fig:SourceSpaceDistribution}.

In our sample, 10,353 sources are accompanied by LOS velocity errors, with 72\% of them having errors $\leq$ 20\,km\,s$^{-1}$, and 66\% having errors $\leq$ 10\,km\,s$^{-1}$.

\subsection{Coordinate and velocity transformations}
\label{sec:Coordinate Transformation of Position and Velocity}

\subsubsection{Coordinate transformation}
\label{sec:Coordinate transformation}

In this study, we adopt the coordinates of the M31 center as ($\alpha_{\rm M31}$, $\delta_{\rm M31}$) = (10.6847083\degr\ , 41.26875\degr\ ) \citep[][]{deVaucouleurs1958}, the position angle and inclination angle of the M31 disk as 38\degr\ \citep[][]{Kent1989} and 77\degr\ \citep[][]{Walterbos1987}, respectively, and the distance of M31 as $d_{\rm M31}=780$\,kpc \citep[][]{Holland1998}. To simplify analysis, we establish a Cartesian coordinate system projected in M31 disk, with the X-axis coinciding with the M31 major axis in the northeast direction and the Y-axis coinciding with the M31 minor axis in the northwest direction. 

For each source, we are able to obtain its major axis projection distance ($X$) and minor axis projection distance ($Y$). However, determining its radii, which is the distance between the object and the center of M31, is challenging as the actual distance from the Sun is unknown. Therefore, we use different methods to calculate radii values for objects in the disk and in the bulge/halo of M31. For disk objects which are assumed to be concentrated on the disk plane, we use their projected distance in the disk. This gives us elliptical contours of radii ($R = \sqrt{X^2+(Y/\cos{77\degr})^2}$). For sources in the halo/bulge which are generally spherically distributed, we use their two-dimensional (2D) projected distance in the sky. This leads to round contours of radii ($r_p = \sqrt{X^2+Y^2}$). In the current work, we use a combination of these radii, denoted by $\mathcal{R}$, as shown in Fig.~\ref{fig:CumulativeMass}. Furthermore, the symbol $r$ is utilized to represent the three-dimensional (3D) radii.

\subsubsection{Velocity Transformation}
\label{sec:Velocity Transformation}

To transform velocity, we first convert the LOS velocity of each object into the Galactocentric coordinate system, and then into the M31-centric coordinate system. This allows us to decompose the peculiar velocity ($v_{\text{pec}}$) of an object into three components, as:
\begin{equation}
	v_{\text{pec}} = v_{\star \rightarrow \odot} + v_{\odot \rightarrow {\rm MW}} + v_{{\rm MW} \rightarrow {\rm M31}}
\end{equation}
where $v_{\star \rightarrow \odot}$ is the motion of the object relative to the Sun, which is the LOS velocity that can be derived from its spectrum, $v_{\odot \rightarrow {\rm MW}}$ is the motion of the Sun relative to the Galactic center, and $v_{{\rm MW} \rightarrow {\rm M31}}$ is the motion of the Galactic center relative to the center of M31.

For a given object, we first calculate its motion relative to the Galactic centre, i.e., $v_{\star \rightarrow {\rm MW}}$. In the current work, we adopt the Galactic rotation velocity at the position of the Sun as 239.3\,km\,s$^{-1}$ \citep[][]{McMillan2011} and the Solar peculiar motion as ($U_\odot$, $V_\odot$, $W_\odot$) = (11.1, 12.24, 7.25)\,km\,s$^{-1}$  \citep{Schonrich2010}. We then have
\begin{equation}
		v_{\star \rightarrow {\rm MW}} = v_{\star \rightarrow \odot} + (239.3 + V_\odot) \sin l \cos b + U_\odot \cos l \cos b + W_\odot \sin b,
\end{equation}
where $l$ and $b$ are longitude and latitude of the object in the Galactic coordinates.

We then compute the motion of the object relative to the center of M31, i.e., $v_{\text{pec}}$. To accomplish this, we decompose the velocity of M31 relative to the Galactic center in the Galactocentric frame into the Galactocentric radial velocity ($v_{\rm M31r}$) and Galactocentric transverse velocity ($v_{\rm M31t}$). We adopt ($v_{\rm M31r}$, $v_{\rm M31t}$) = ($-$109, 17)\,km\,s$^{-1}$ \citep[][]{vanderMarel2008,vanderMarel2012} and a position angle of $\theta _{\rm M31t} = 287$\degr\ \citep[][]{vanderMarel2012}, then $v_{\text{pec}}$ is given by
\begin{equation}
	v_{\text{pec}}= v_{\star \rightarrow _{\rm MW}} - v_{\rm M31r} \cos \varrho  + v_{\rm M31t} \sin \varrho  \cos (\phi - \theta _{\rm M31t}),
\end{equation}
where $\varrho $ is the angular separation between the object and the center of M31, and $\phi$ is the position angle of the object with respect to the M31 center. 

\section{Kinematics modeling: the disk of M31}
\label{sec:Kinematics Modelling: Disk}

\begin{figure}
	\includegraphics[width=\columnwidth]{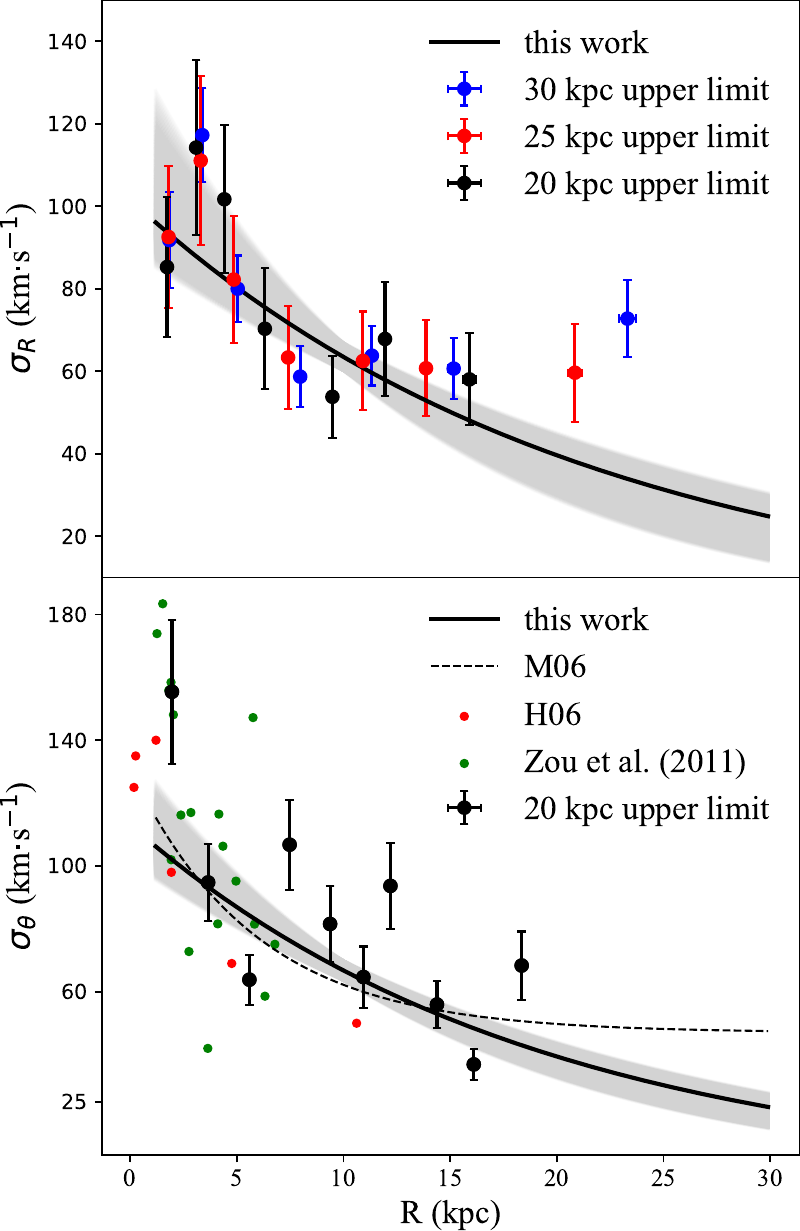}
    \caption{Dispersions of the radial velocity (upper panel) for the selected objects close to the minor axis and the tangential velocity (bottom panel) for the selected objects close to the major axis. The measured dispersion values for individual radius bins (using 20\,kpc as the disk limit) are represented by black dots with error bars. The best-fit curves are shown as black lines, and the 1$\sigma$ regions of the best fits are depicted with light grey shadings. In the upper panel, additional blue and red dots with error bars represent disk limits of 25 and 30,kpc. The black dashed curve, black dots and green dots in the bottom panel show the results from M06, H06, and \citet{Zou2011}, respectively.} 
    \label{fig:DispersionFitting}
\end{figure}

\begin{figure}
\centering
	\includegraphics[width=\columnwidth]{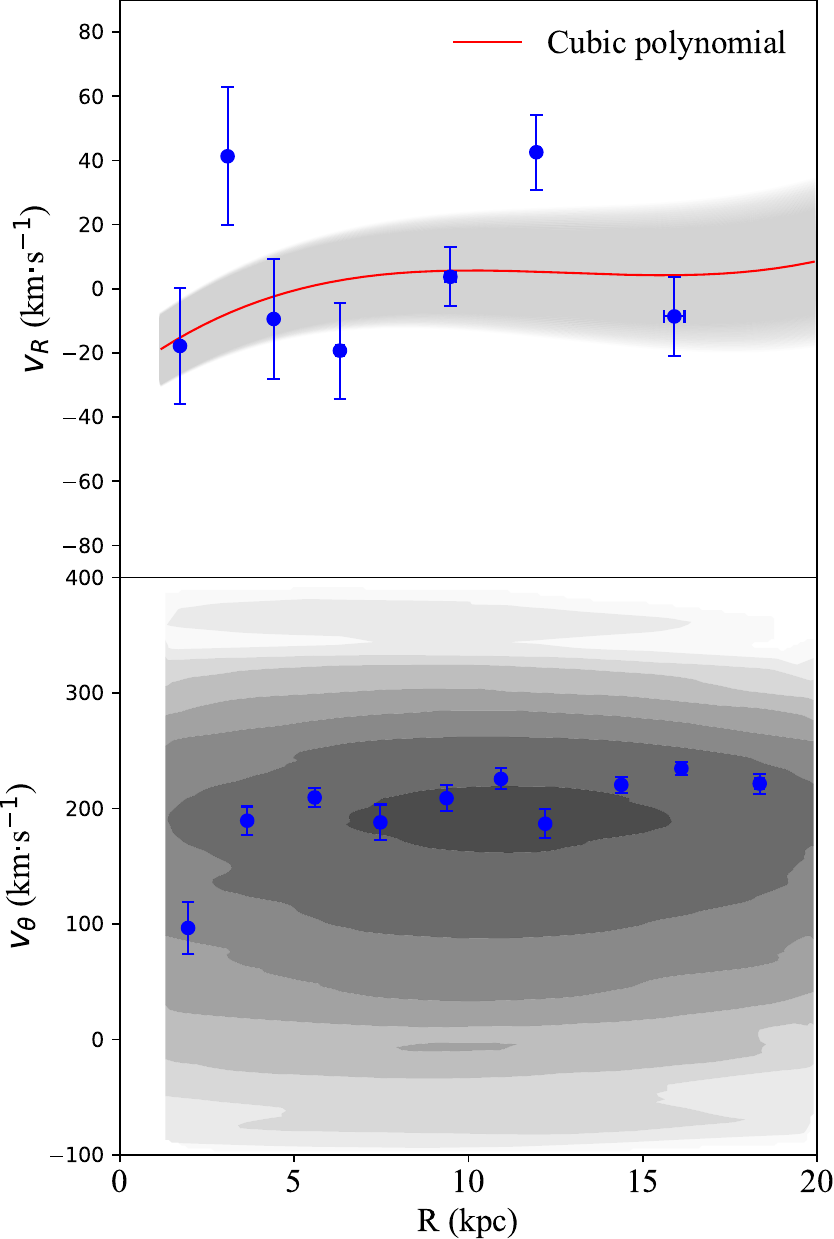}
	\caption{{\it Upper panel}: Mean values of the radial velocities for the selected objects close to the minor axis (blue dots). The red line and light grey shading represent the best cubic polynomial fit curve and the 1$\sigma$ region error respectively, used to infer the radial velocity of all the disk objects. {\it Bottom panel}: Mean values of the tangential velocities for the selected objects close to the major axis (blue dots) and the density distribution of estimated tangential velocities for all the disk sources of M31 (grey scales). The darker the color, the greater the density. }
    \label{fig:MeanVelocity}
\end{figure}

To properly discuss the kinematics of the disk of M31, it is crucial to first consider the range of the disk and the relevant tracers. The commonly accepted range where the disk dominates is between 1.2 and 9\,kpc \citep[][]{Courteau2011}. Rotation curves based on H~{\sc i} gas have been observed at much farther distances, such as 38\,kpc \citep[][]{Chemin2009}. Moreover, extensive kinematic surveys \citep[e.g.][]{Ibata2005} and wide-field imaging surveys \citep[e.g.][]{Bernard2015} have shown that M31 has an extended disk, spanning radii from 15\,kpc to 40\,kpc and exhibiting extended disk-like substructures at around 60\,kpc. In our study, we have explored various disk limits, favoring a more conservative approach to minimize contamination from halo objects and obtain a relatively pure sample of the disk. We plot the radial velocity dispersion values for objects located close to the minor axis of M31, using radii limits of 20, 25, and 30\,kpc, respectively, in the upper panel of Fig.~\ref{fig:DispersionFitting}. Within the radius bins of 20\,kpc, we observe a decrease in the radial velocity dispersion values as the radius increases. However, when we consider disk limits of 30 and 25\,kpc, we observe an increase in radial velocity dispersion beyond 20\,kpc, indicating contaminations from halo objects. Among the objects in our sample, 89\% of the H~{\sc ii} regions and candidates, which are believed to be young disk objects, are located within the range of 1.2--20\,kpc. H~{\sc ii} regions and candidates beyond this range mainly come from the M06 catalogue and are considered candidates rather than confirmed H~{\sc ii} regions. Therefore, we adopt a maximum radius limit of 20\,kpc for the kinematic analysis of the M31 disk. For our analysis of the M31 disk, we have focused on tracers such as stars (including OB stars, supergiants, and other types) and emission-line objects (including PNe, H~{\sc ii} regions, and unclassified objects) located within the selected disk radii range, i.e., from 1.2 to 20\,kpc.

For the subsequent kinematic analysis, we utilize the cylindrical coordinate system centered on the centre of M31, which is similar to the one in Section~\ref{sec:Coordinate transformation} but with slight differences. For disk objects, we disregard their vertical velocity component in the $Z$-direction, which is perpendicular to the M31 disk. We measure the angle coordinates $\theta$ in a counterclockwise direction from the negative semi-axis of the X-axis and the radial coordinate $R$ increasing outward. The positive direction of the tangential velocity $v_\theta$ is opposite to the direction of $\theta$ increase. By adopting this convention, we ensure that the disk sources have positive circular velocities. The positive direction of the radial velocity $v_R$ points away from the center of M31. We can then decompose $v_{\rm pec}$ into two components: $v_R$ and $v_\theta$, given by,
\begin{equation}
	v_{\text{pec}} = -v_R\sin \theta \sin i + v_\theta \cos \theta \sin i,
	\label{eq:vpeccomponent}
\end{equation}
where $i$ is the inclination angle of the disk of M31.

The tangential velocity $v_\theta$ is not equivalent to the circular velocity $v_c$ due to the non-negligible effect of asymmetric drift ($v_a$), which arises from the heating effect of the disk. We assume the disk is in a steady state and ignore the variations of covariance $\overline{v_Rv_z}$  with Z. The density of tracers in the disk and their radial velocity dispersion, $\sigma_R$, both decrease exponentially with $R$, and the scale lengths for the decrease in number density and radial velocity dispersion are $L_d$ and $L_R$, respectively. The asymmetric drift is then given by \citep[e.g.][]{Binney2008} 
\begin{equation}
	v_a (R) = {\sigma _R^2(R)\over {2v_c(R)}} [{{\sigma _\theta^2(R)}\over {\sigma_R^2(R)}}-1+R({1\over L_d} + {2\over L_{R}})],
	\label{eq:asymmetricdrift}
\end{equation}
where $\sigma_\theta$ is the tangential velocity dispersions. In this work we adopt a value of $L_d = 5.3\pm 0.5$\,kpc based on the work of \citep[][]{Courteau2011}. For the rest of the parameters, we will derive them from our sample as some are not available in the literature at present, and some are accompanied by considerable uncertainty.

\subsection{Radial and tangential velocity dispersions}
\label{sec:Radial Velocity Dispersion}

The radial velocity dispersion $\sigma_R(R)$ of the disk source in M31 is still uncertain. It is very difficult to measure the radial velocity $v_R$ of an M31 disk source directly. However, we can approximate the LOS velocity and dispersion of the source located on the minor axis of the M31 disk as a radial velocity. We select objects located in a fan-shaped region near the minor axis of the M31 disk, as given by: $|X/R| < 0.1$. For these sources, their tangential velocity $v_{\theta}$ is perpendicular to our LOS, so the contribution of the LOS velocity to $v_{\theta}$ can be neglected. The value of the radial velocity $v_R$ of these objects can be calculated by,
\begin{equation}
	v_R = -v_{\text{pec}}/ (\sin \theta \sin i ).
	\label{eq:v_r}
\end{equation}
The selected tracers are divided into seven radius bins, each containing more than 35 sources. The mean and dispersion uncertainties of the individual bins are estimated from 300 bootstrapping samples. The top panel of Fig.~\ref{fig:DispersionFitting} show the dispersion values for these seven bins.

As in the case of the Galactic disk \citep[e.g.][]{Huang2016}, we assume that the radial velocity dispersions $\sigma_R$ of the objects in the disk of M31 decrease exponentially with respect to the radii $R$, given by,
\begin{equation}
	\sigma_R(R) = \sigma_{R_0} \exp ( -{{R-R_0}\over L_{R}}),
	\label{eq:sigma_r}
\end{equation}
where $\sigma_{\text{LOS}_0}$ is the radial velocity dispersion at the reference position $R_0$. In the current work, we adopt $R_{0}=10$\,kpc considering the size of the disk of M31. We use Monte Carlo Markov Chain (MCMC) analysis to extract the best-fit values for $\sigma_{R_0}$ and $L_R$. As a result, we obtain: $\sigma_{R_0} = 62.7^{+3.4}_{-3.4}$\,km\,s$^{-1}$ and $L_{R} = 19.8^{+6.5}_{-4.0}$\,kpc. The best-fit curve is illustrated in top panel of Fig.~\ref{fig:DispersionFitting}. 

In a similar manner, for the tangential velocity $v_{\theta}$, we select objects that are located close to the major axis of the disk of M31, where the radial velocity $v_R$ is perpendicular to our LOS. Only objects that meet the criterion $|Y/R|<0.07$ are chosen, and their tangential velocity $v_{\theta}$ can be calculated by,
\begin{equation}
	v_\theta = v_{\text{pec}} /(\cos \theta \sin i ).
	\label{eq:v_theta}
\end{equation}
We group the selected objects into ten radius bins, with tracer size of each bin larger than 45. The bottom panel of Fig.~\ref{fig:DispersionFitting} presents the tangential velocity dispersion values for each bin. We have compared our results with those obtained from previous studies, including M06, H06, and \citet{Zou2011}, which are over-plotted in the Fig.~\ref{fig:DispersionFitting}. The results demonstrate a good agreement with the previous studies. We used an exponential function similar to equation~(\ref{eq:sigma_r}) to model the tangential velocity dispersion, given by,
\begin{equation}
	\sigma_{\theta}(R) = \sigma_{\theta_0} \exp ( -{{R-R_0}\over L_{\theta}}),
	\label{eq:sigma_theta}
\end{equation}
where $\sigma_{\theta_0}$ is the tangential velocity dispersion at $R=R_0$ and $L_{\theta}$ the scale length of the tangential velocity dispersion. Similarly, we have applied the MCMC analysis to estimate the best-fit parameters, which yields values of $\sigma_{\theta_0} = 69.6^{+3.3}_{-3.3}$\,km\,s$^{-1}$ and $L_{\theta} = 21.0^{+5.2}_{-3.5}$\,kpc. The corresponding best-fit curve is shown in bottom panel of Fig.~\ref{fig:DispersionFitting}.

\subsection{Tangential velocity and asymmetric drift correction}

In order to determine the circular velocity of objects in M31 disk, we need to make the asymmetric drift corrections to their tangential velocity. We previously computed the tangential velocities of a small subset of the disk objects (508 out of 5261 tracers) that are located near the major axis of the M31 disk. Now, we aim to estimate the tangential velocities of all the selected disk objects. To do so, we assume a radial velocity function for all disk objects. Ideally, the radial velocities of objects in a stable, symmetric disk would be zero. However, as depicted in upper panel of Fig.~\ref{fig:MeanVelocity}, the mean radial motion $\overline{v_R}$ oscillates with $R$, much like what has been observed in the Milky Way \citep[][]{Williams2013,Huang2016,Wang2023}. The mean radial velocity varies around zero, indicating either an elliptical mean orbit of the disk objects or the process of radial migration. To smooth out this oscillation relation, we apply a cubic polynomial fit, which is also shown in Fig.~\ref{fig:MeanVelocity}. Based on the fitted relation, we calculate the radial velocity of all the disk objects, then the tangential velocity of each object can be derived using equation~(\ref{eq:vpeccomponent}). The distribution of tangential velocities calculated for all disk sources in M31 is shown in the lower panel of Fig.~\ref{fig:MeanVelocity}. In the diagram we also show the mean values of the tangential velocities for the objects located close to the major axis of the disk of M31. The tangential velocities vary with radius in a manner very similar to that observed for sources located near the major axis of M31, demonstrating the robustness of our results.

With the tangential velocities of all the selected disk objects and all parameters except for the circular velocity to calculate the asymmetric drift corrections, we can now determine the circular velocities of the individual objects in the disk of M31. This is achieved by combining equation~(\ref{eq:asymmetricdrift}) with the relation $v_c = v_{\theta} + v_a$. Tracers located near the minor axis ($|X/R| < 0.1$) are excluded from the analysis, as their peculiar velocities have minimal contribution to the circular velocities, leading to a larger uncertainty. To account for the errors we use the Monte Carlo (MC) method.

\section{Kinematics Modelling: the Bulge and Halo of M31}
\label{sec:Kinematics Modelling: Bulge and Halo}

We first select objects that belong to the bulge and halo of M31. According to Section~\ref{sec:Kinematics Modelling: Disk}, we set the radii range of the bulge as $R < 1.2$\,kpc, and that of the halo as $R > 30$\,kpc. Halo stars from DESI observations, PNe, unclassified emission-line objects, and clusters that fall within these radius ranges are selected. Due to the different methods used to calculate the radii of disk and halo objects, there is overlap between the disk and the halo in the resulting rotation curve (see Fig.~\ref{fig:RCP}). This overlap will be utilized to constrain the halo anisotropy parameter. In the subsequent calculations of this section, we will employ the 3D radius ($r$) and the 2D projected radius ($r_p$) in the sky (see Section~\ref{sec:Coordinate transformation}).

Unlike the disk, both the bulge and the halo of M31 are assumed to be spherical gravitational potential systems. To model the kinematics of these systems, we use the Jeans equation given by \citep{Jeans1919,Binney2008}, 
\begin{equation}
	{{\text{d} (\rho {\sigma_r^2})}\over {\text{d} r}}+ 2{{\beta}\over r}\rho {\sigma_r^2} = -\rho {{\text{d} \Phi}\over {\text{d} r}},
	\label{eq:HaloJeansEquation}
\end{equation}
where $\rho$ is the number density, $\sigma_r$ is the radial velocity dispersion, $\beta$ is the anisotropy parameter, and $\Phi$ is the total gravitational potential. The circular velocity is associated with the gravitational potential by,
\begin{equation}
v_c^2= r{{d\Phi}\over{d r}}.
\label{eq:v_candPhi}
\end{equation}

The number density $\rho$ of the bulge and halo of M31, as utilized in this study, is based on the results of \citet{Courteau2011}. The bulge is characterized by a projected profile that can be described by the S\'{e}rsic form, with a S\'{e}rsic shape index of $n = 2.2\pm0.3$, a S\'{e}rsic shape parameter $b_n = 1.9992n-0.3271$, and an effective radius $R_e = 1.0 \pm 0.2$\,kpc.
On the other hand, the halo exhibits a power law profile in both 2D and 3D. This can be expressed as a 2D surface density profile, $\Sigma = \Sigma_0 r_p^{\alpha_{\text{2D}}}$, and a 3D density profile, $\rho = \rho_0 r^{\alpha_{\text{3D}}}$, where $\Sigma_0$ and $\rho_0$ represent the 2D and 3D densities at 1\,kpc, respectively. The power law indices are chosen as $\alpha_{\text{2D}} = -2.5\pm 0.2$ and $\alpha_{\text{3D}} = -3.5\pm 0.2$ \citep{Courteau2011}.

Following the works of \citet{Veljanoski2014} and \citet{Gilbert2018}, we adopt a power law for the radial velocity dispersion, $\sigma_r$, such that $\sigma_r \propto {r }^\gamma$.

Now the circular velocities of the bulge and halo are respectively given by, 
\begin{equation}
	v_{c\text{bulge}}^2 = - \sigma_r^2(2\gamma - {b_n \over n}({r\over R_e})^{1/n} + 2\beta_{\text{bulge}}),
	\label{eq:bulgerotation}
\end{equation}
and
\begin{equation}
	v_{c\text{halo}}^2 = - \sigma_r^2(2\gamma  + \alpha_{\text{3D}} + 2\beta_{\text{halo}}).
	\label{eq:halorotation}
\end{equation}
However, since we are unable to directly measure the actual distance and 3D velocity of individual objects in the M31 halo, we can only obtain projected 2D kinematic parameters from observations. In this study, we provide the deprojected circular velocities of the halo as a replacement for equation~(\ref{eq:halorotation}), given by,
\begin{equation}
	\begin{aligned}
	\sigma_{\text{LOS}}^2(r_p) = &{{2 \times \rho_0 /\Sigma_0}\over {-2\gamma-\alpha_{\text{3D}}-2\beta_{\text{halo}}}} r_p^{-\alpha_{\text{2D}}} \times \\
	&\int_{r_p}^{\infty} (1-\beta_{\text{halo}}{{r_p^2}\over {r^2}}) r^{\alpha_{\text{3D}}+1}v_{c\text{halo}}^2 {{dr}\over{\sqrt{r^2-r_p^2}}}.
	\end{aligned}
	\label{eq:sigmaLOSsigma}
\end{equation}
where $\frac{\rho_0}{\Sigma_0}=0.57$\,kpc$^{-1}$ is the ratio of number density to surface density at 1\,kpc. The derivation of this equation is described in Appendix~\ref{sec:DeprojectedCircularVelocity}. Previous literature does not offer comprehensive 3D information on the circular velocities of the bulge, making it difficult to establish a corresponding equation. Therefore, we continue to use equation~(\ref{eq:bulgerotation}) with 2D parameters to approximate the 3D circular velocities of the bulge. Within the bulge region, there are only two available measurements of circular velocity, and their impact on the virial mass is minimal. Furthermore, the circular velocity values in the bulge are primarily influenced by $\beta_{\text{bulge}}$. Instead of opting for more complex models that incorporate missing information, substituting the 2D scenario is a preferable choice. %Similar experiments have been conducted on the halo, and it has been found that approximating it with a 2D parameterization results in a slightly shallower gradient compared to using equation~(\ref{eq:sigmaLOSsigma}) in the rotation curve of the halo. This introduces a small effect on the rotation curve values, typically within 15 km\,s$^{-1}$, which is smaller than the systematic error of the circular velocity measurements.}

\begin{figure}
	\includegraphics[width=\columnwidth]{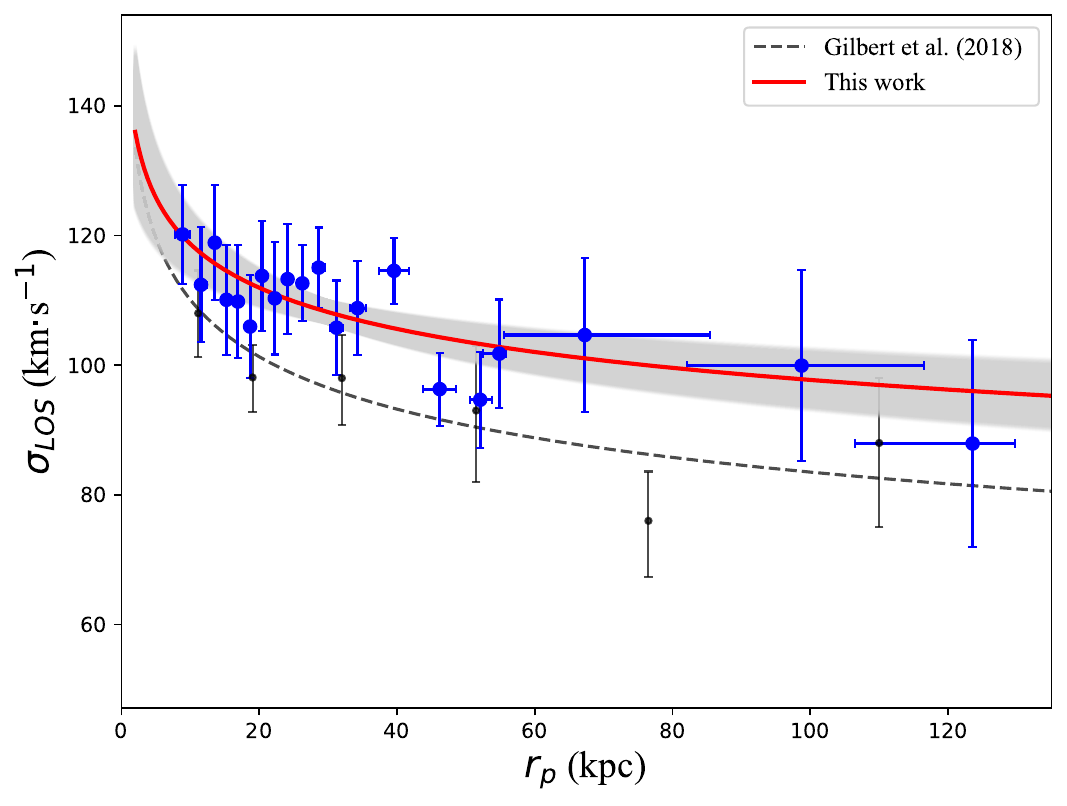}
    \caption{The relationship between $\sigma_{\text{LOS}}$ and $r_p$ for the halo of M31. The blue dots correspond to the $\sigma_{\text{LOS}}$ values for each of the radius bins. The red curve and the grey shading represent the best-fit curve and the 1$\sigma$ region of the fit, respectively. The black dashed line and black points with errorbars denotes the velocity dispersion from \citet{Gilbert2018}.}
    \label{fig:HaloDispersionFitting}
\end{figure}

To calculate $\sigma_{\text{LOS}}$ for the halo of M31, we divide the halo objects into 20 radial bins. In the case of the inner 15 bins, as the radius increases, the sky area covered by each radial bin widens progressively, resulting in additional dispersion beyond the intrinsic value. To address this, we partition the objects within each bin into four quadrants using the $X$ and $Y$ axes of M31. Subsequently, we calculate the LOS velocity dispersion for each quadrant and derive a weighted average as the velocity dispersion value for the bin, considering the number of objects in each quadrant. For the outer 5 bins, we do not differentiate between quadrants due to the limited number of objects in our sample and calculate the LOS velocity dispersion directly. Fig.~\ref{fig:HaloDispersionFitting} displays $\sigma_{\text{LOS}}$ of the halo objects as a function of $r_p$, with previous results from \citet{Gilbert2018} over-plotted. Our results are generally consistent with \citet{Gilbert2018}, though slightly higher. We fit $\sigma_{\text{LOS}}$ as a power law \citep[][]{Veljanoski2014,Gilbert2018}, given by, \begin{equation}
	\sigma_{\text{LOS}} = \sigma_{\text{LOS}_0} ({r_p \over r_{p_0}})^\gamma,
	\label{eq:halordispersion}
\end{equation}
where we keep $r_{p_0}$ fixed at 30\,kpc and use MCMC analysis to determine the parameters $\sigma_{\text{LOS}_0}$ and $\gamma$. Our analysis yields $\sigma_{\text{LOS}_0}= 107.88 ^{+1.77}_{-1.74}$\,km\,s$^{-1}$ and $\gamma = -0.09 \pm0.03 $. The best-fit curve is displayed in Fig.~\ref{fig:HaloDispersionFitting}.

To obtain the $\sigma_{\text{LOS}}$ of the bulge, we separate the bulge objects into two radius bins and compute the radii and LOS velocity dispersions for each bin. However, bulge velocity dispersion is still assumed in power law form as equation~(\ref{eq:halordispersion}). As there are only two $\sigma_{\text{LOS}}$ measurements within $0<R<1.2$\,kpc for the bulge, we have not attempted to fit a velocity dispersion relation. $\gamma = -0.09 \pm0.03 $ obtained from the halo fits is adopted for both the bulge and halo analysis.

Finally, given the lack of knowledge about the anisotropic parameter of the bulge and halo of M31 ($\beta_{\text{bulge}}$ and $\beta_{\text{halo}}$), we treat them as free kinematic parameters in conjunction with the dynamical parameters determined in Section~\ref{sec:Dynamics modelling: gravitational potential}. However, as observed in the Milky Way, the anisotropy parameters exhibit spatial variations. For the bulge and halo of M31, accounting for the influence of radius on $\beta$ presents challenges, so we only constrain an spatially averaged value for these parameters.

\section{The rotation curve of M31}
\label{sec:result and discussion}

\begin{figure*}
	\includegraphics[width=1.6\columnwidth]{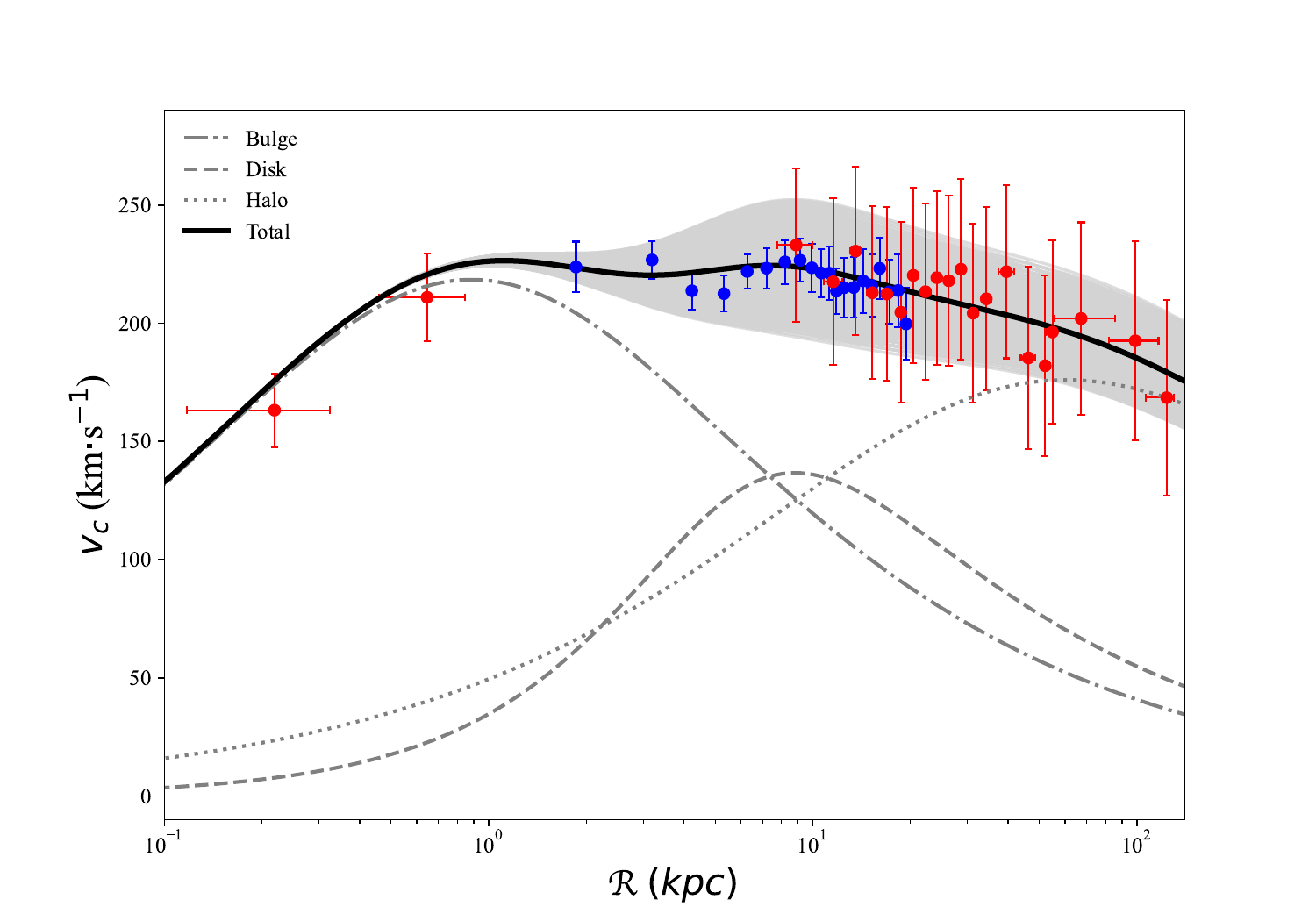}
    \caption{Final combined 3-part rotation curve of M31, comprising the disk (blue dots), bulge and halo (red dots). The different line styles, as indicated in the top-left corner of the diagram, represent the best-fit rotation curve contributions to the individual components of M31. The black line represents the sum of contributions from all the mass components, while the grey shading corresponds to the 1$\sigma$ region of the fit.}
    \label{fig:RCP}
\end{figure*}

\begin{table}
	\centering
	\caption{The rotation curve of M31.}
	\label{tab:RotCurve}
	\begin{tabular}{cccccc}
	\hline
	$\mathcal{R}$  & $\mathcal{R}$ error  & $v_c$ & $v_c$ error& fitted $v_c$  \\
	(kpc) & (kpc) & (km\,s$^{-1}$)  & (km\,s$^{-1}$)  & (km\,s$^{-1}$)  \\
	\hline
	Bulge &  & & &  \\
	\hline
	0.22&0.1&163.24&15.61&175.78\\
	0.65&0.19&210.95&18.56&220.61\\
	\hline
	Disk &  & & &  \\
	\hline
	1.86&0.02&223.8&10.72&223.43\\
	3.19&0.02&226.79&7.96&220.33\\
	4.23&0.02&213.74&8.17&221.38\\
	5.32&0.02&212.54&7.55&223.04\\
	6.27&0.02&221.94&7.2&224.04\\
	7.2&0.02&223.3&8.49&224.43\\
	8.2&0.02&225.97&9.28&224.28\\
	9.14&0.02&226.63&9.07&223.75\\
	9.93&0.02&223.45&10.29&223.1\\
	10.62&0.01&221.25&10.24&222.44\\
	11.25&0.01&221.17&11.16&221.79\\
	11.83&0.01&213.58&9.75&221.17\\
	12.48&0.01&215.04&12.67&220.44\\
	13.37&0.02&215.12&12.82&219.48\\
	14.29&0.02&217.92&13.52&218.48\\
	15.2&0.02&215.97&13.09&217.54\\
	16.08&0.02&223.22&13.07&216.67\\
	17.21&0.02&213.32&13.44&215.61\\
	18.31&0.03&213.85&15.37&214.64\\
	19.38&0.03&199.73&15.19&213.76\\
	\hline
	Halo &  & & &  \\
	\hline
	8.88&1.08&233.09&32.4&223.92\\
	11.55&0.77&217.54&35.26&221.46\\
	13.55&0.57&230.53&35.63&219.28\\
	15.25&0.54&212.91&36.6&217.49\\
	16.91&0.62&212.36&36.67&215.88\\
	18.68&0.54&204.64&38.1&214.33\\
	20.41&0.54&220.27&37.18&212.98\\
	22.26&0.62&213.35&37.17&211.68\\
	24.12&0.67&219.28&36.73&210.51\\
	26.26&0.76&217.99&35.88&209.29\\
	28.63&0.87&222.84&38.16&208.08\\
	31.23&0.88&204.28&37.89&206.88\\
	34.28&1.19&210.33&38.76&205.58\\
	39.56&2.21&221.84&36.68&203.53\\
	46.21&2.42&185.33&38.68&201.19\\
	52.08&1.63&182.05&38.15&199.23\\
	54.85&1.67&196.28&38.81&198.33\\
	67.26&14.95&202.02&40.67&194.47\\
	98.74&17.19&192.59&42.23&185.55\\
	123.56&11.6&168.53&41.34&179.31\\
	\hline
	\end{tabular}
\end{table}

\begin{table*}
	\centering
	\caption{Kinematic parameters used to calculate the rotation curve.}
	\label{tab:kinematicpara}
	\begin{tabular}{lcccr} 
		\hline
		Component & Parameter& Description & Value  & Note \\
		\hline
		Bulge & $\gamma$  & Velocity dispersion power law index & $-0.09\pm$0.03& fixed\\
		~     & $n$ &  S\'{e}rsic shape index of the number density distribution & $2.2\pm0.3$ & fixed\\
		~     & $R_e$ & S\'{e}rsic effective radius of the number density distribution  & $1.0 \pm 0.2$\,kpc& fixed\\
		~     & $\beta_{\text{bulge}}$ & Anisotropy parameter of the bulge & $ 0.07^{+0.13}_{-0.29}$ & fitted\\
		\hline
		Disk  & $R_0$ & Reference radius & 10\,kpc & fixed\\ 
		~     & $\sigma_{R_0}$  & Radial velocity dispersion at $R_0$ & $62.7^{+3.4}_{-3.4}$\,km\,s$^{-1}$ & fitted\\
		~     & $L_{R}$ & Scale length of radial velocity dispersion & $19.8^{+6.5}_{-4.0}$\,kpc & fitted\\		
		~     & $\sigma_{\theta_0}$ & Tangent velocity dispersion at $R_0$ &$69.6^{+3.3}_{-3.3}$\,km\,s$^{-1}$& fitted\\
		~     & $L_{\theta}$ & Scale length of tangent velocity dispersion & $21.0^{+5.2}_{-3.5}$\,kpc & fitted\\
		~     & $L_{d}$  &  Scale length of the number density & $5.3\pm0.5$\,kpc&fixed\\
		\hline
		Stellar halo  & $\gamma$  & Velocity dispersion power law index & $-0.09\pm$0.03& fitted\\
		~     & $\alpha_{\text{2D}}$  & Number density power law index & $-2.5\pm0.2$ & fixed\\
		~     & $\alpha_{\text{3D}}$  & Number density power law index & $-3.5\pm0.2$ & fixed\\
		~     & $\beta_{\text{halo}}$ & Anisotropy parameter of the halo & $ -0.54 ^{+0.38}_{-1.29}$ & fitted\\
		\hline
	\end{tabular}
\end{table*}

Using the most extensive collection of velocity measurements of M31 objects to date, we have generated a rotation curve for M31 that encompasses all of its components, including the bulge, disk, and halo. Hereafter, we adopt the terminology of ``rotation velocity'' and ``rotation curve'' rather than ``circular velocity'' to maintain consistency with prior studies. However, it is important to note that these concepts are only equivalent in the context of a cold disk. The rotation velocity of stars is typically lower than the circular velocity, and in some literature, it is even assumed to be effectively zero in the halo. The resulting rotation curve extends out to a distance of $\sim$125\,kpc, which is presented in Fig.~\ref{fig:RCP}. Table~\ref{tab:RotCurve} displays the values of our calculated rotation curve. Table~\ref{tab:kinematicpara} presents a summary of all the kinematic parameters utilized in the calculation of the rotation curve for the current study.

\subsection{Rotation curve of the disk of M31}

\begin{figure}
	\includegraphics[width=\columnwidth]{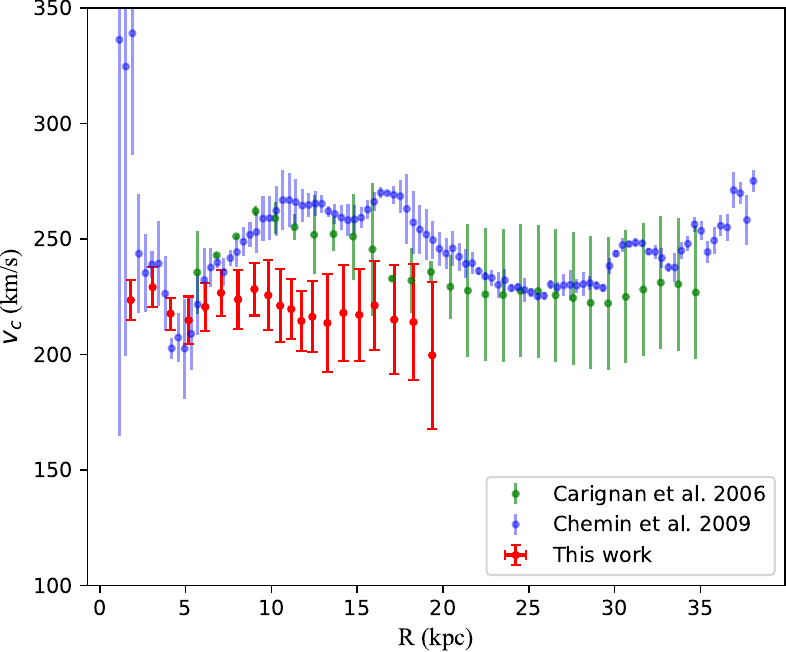}
	\caption{Rotation curve of the disk of M31. The red dots represent the results from this study, and the blue and green dots are the previous M31 rotation curves obtained from the H~{\sc i} studies.}
    \label{fig:DiskRotatingCurve}
\end{figure}

The resulting rotation curve of the disk of M31, along with a comparison to previous H~{\sc i} studies \citep{Carignan2006, Chemin2009}, is shown in Fig.~\ref{fig:DiskRotatingCurve}. The disk objects are divided into 20 radius bins, each containing over 220 objects. The typical rotation velocity in the disk of M31 is about 220\,km\,s$^{-1}$. Our results generally agree with the previous rotation curves obtained from H~{\sc i} observations. However, for regions at radius $R$ = 7 -- 18\,kpc, our estimates for the rotation curve are slightly lower than those from the H~{\sc i} studies. Since the difference is relatively small and still falls within the 3$\sigma$ range, it is probable that the dissimilarity is caused by systematic differences in the tracers and methods used to measure the rotation curve.

Our calculated rotation velocities in the disk of M31 have a typical uncertainty of about 17\,km\,s$^{-1}$. The most significant contributor to this uncertainty is the tangential velocity error. Since we cannot directly measure the tangential velocities of the individual objects in the disk, we have assumed a radial velocity and radius relationship to derive the tangential velocities. Fig.~\ref{fig:MeanVelocity} shows that the uncertainties of the fitted radial velocities can range from 9 to 30\,km\,s$^{-1}$, leading to an error of around 5 to 38\,km\,s$^{-1}$ in the obtained rotation velocity. Typically, this results in an error of approximately 16\,km\,s$^{-1}$. Aside from this, there are other sources of uncertainty, including errors in the parameters used. For instance, the scale length of number density $L_d$ can have a maximum value of 6.5\,kpc \citep{Kormendy1999}. Using a different value of $L_d$  would result in a systematic rotation velocity error of about 5\,km\,s$^{-1}$. Finally, the uncertainty of rotation curve caused by the error of the scale length of radial velocity dispersion $L_R$ is about 2\,km\,s$^{-1}$, while that from the error of the scale length of tangential velocity dispersion $L_{\theta}$ is about 1\,km\,s$^{-1}$.

%\textbf{Here we discuss the impact of different upper limits of disk. Although the large difference in $L_R$ affects the asymmetric drft correction, the impact on the final rotation curve is within 5km/s in 1.2-12\,kpc. Rotation curve in 25 and 30\,kpc upper limits have the same trend as the upper limit of 20kpc. The difference in rotation curves caused by the upper limits begins to increase after 12\,kpc. The rotation velocity in the 30\,kpc case is greater than that in the 20kpc case, and reaches the maximum 15\,km\,s$^{-1}$ at 20kpc.}

\subsection{Rotation curve of the bulge and halo of M31}

The objects in the bulge of M31 has been divided into two radius bins. The rotation velocities are 163.24 and 210.95\,km\,s$^{-1}$ at distances of 0.22 and 0.65\,kpc, respectively. These values are lower than those of the disk. 

The uncertainty of the rotation velocities of the bulge is about 33\,km\,s$^{-1}$. This can be attributed to our lack of understanding of the bulge parameters, which themselves have large errors. The S\'{e}rsic effective radius of the bulge number density distribution $R_e$ has a maximum reported value of 1.93 by \citet{Seigar2008}, which could result in a systematic error of approximately 30\,km\,s$^{-1}$ for the rotation velocity. measurement of the S\'{e}rsic shape index of the bulge number density $n$ has an uncertainty that could lead to a systematic error of approximately 12\,km\,s$^{-1}$ for the rotation velocity. Additionally, for the velocity dispersion power law index $\gamma$, \citet{Gilbert2018} has measured a value of $-0.12$, which could result in a systematic error of about 5\,km\,s$^{-1}$ for the rotation velocity.

Rotation velocities are calculated in 20 radius bins for the halo of M31. Since the way for measuring radii in the disk differs from that used for objects in the halo (as described in Section~\ref{sec:Coordinate transformation}), there are some overlapped bins in the rotation curve. In these overlapped radii, the rotation curve of the disk agrees well as that of the halo, which strengthens the reliability of our results. The rotation curve of the halo exhibits a radial decline, decreasing from around 230\,km\,s$^{-1}$ at a distance of 30\,kpc to approximately 170\,km\,s$^{-1}$ at 125\,kpc.

Equation~(\ref{eq:sigmaLOSsigma}) demonstrates high sensitivity to kinematic parameters, making it crucial to obtain precise values. Error propagation analysis reveals that a mere 0.2 error in $\alpha_{\text{2D}}$ and $\alpha_{\text{3D}}$ can introduce a substantial error of 30\,km\,s$^{-1}$ into the rotation curve. The surface density power law index of the halo, denoted as $\alpha_{\text{2D}}$ and documented in \citet{Courteau2011}, exhibits a considerable range from $-3.5$ to $-2.2$. This wide variation in $\alpha_{\text{2D}}$ significantly impacts the halo's rotation curve. Additionally, the power law index $\gamma$, as reported in \citet{Gilbert2018}, is associated with an error of approximately 18\,km\,s$^{-1}$ in the rotation velocities. Fortunately, due to the degeneracy between these kinematic parameters and $\beta_{\text{halo}}$, the uncertainty in halo rotation velocity is somewhat mitigated. However, it remains imperative to achieve more precise measurements of these kinematic parameters, particularly $\alpha$.

\begin{figure*}
	\includegraphics[width=2\columnwidth]{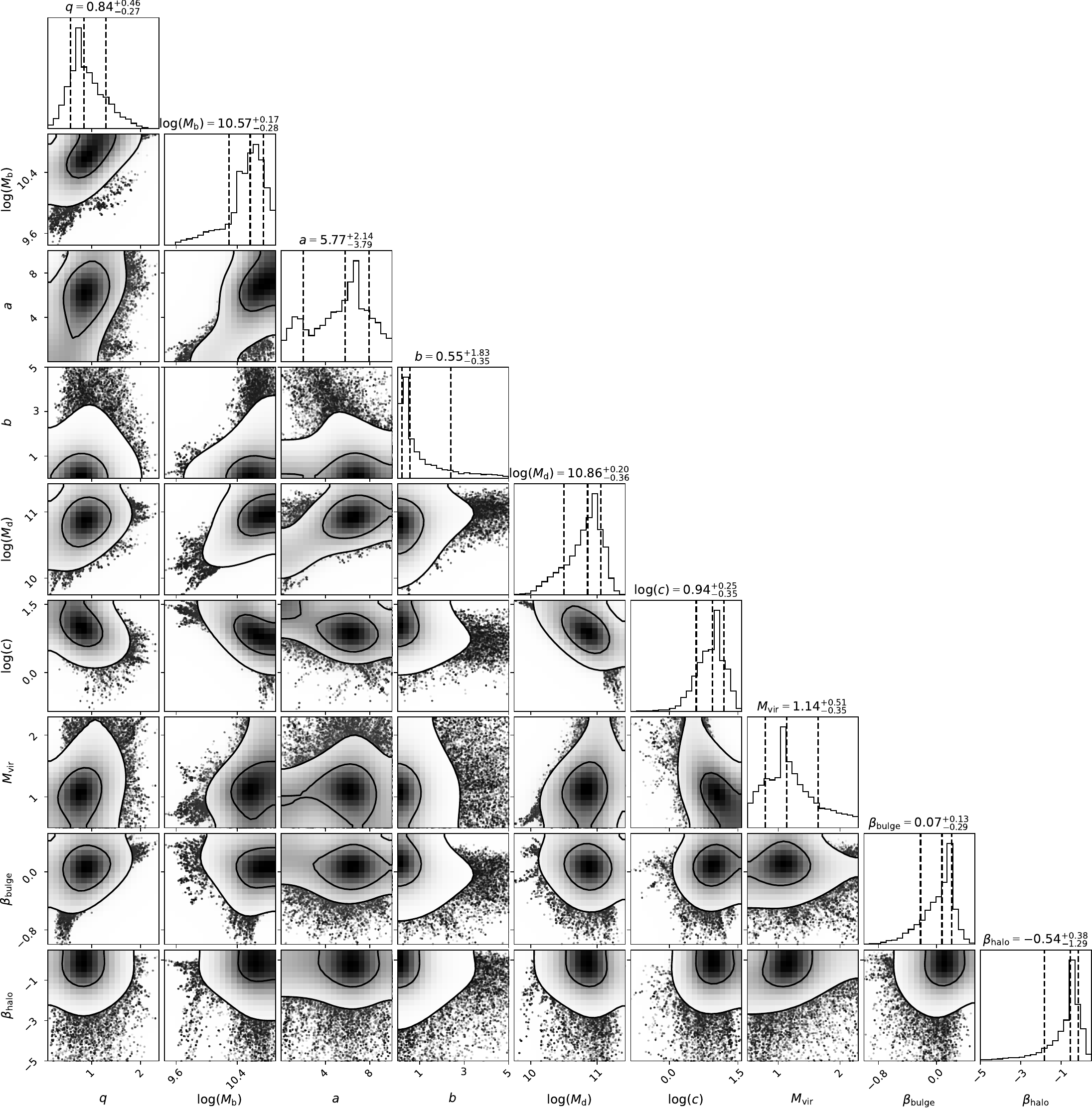}
    \caption{The probability distributions of the free parameters in the MCMC chains that describe the potential of M31. The 2D distribution panels show contours at levels of 0.5, 1, 1.5, and 2$\sigma$. Dashed lines marking the values of the 16th, 50th, and 84th percentiles of the individual parameters can be seen in the 1D distribution panels, with their corresponding values displayed at the top. }
    \label{fig:PotentialCorner}
\end{figure*}

\section{The Mass distribution of M31}
\label{sec:Dynamics modelling: gravitational potential}

\subsection{The gravitational potential model}

We have developed a parameterized potential model to fit the rotation curve of M31 and obtain its mass distribution. The model consists of three components: the bulge, disk, and halo, and the total gravitational potential is obtained by summing these components. 

In the current work, we adopt the potential models proposed by \citet{Hernquist1990} and \citet{Miyamoto1975} for the bulge and disk, respectively. Therefore, the bulge and disk potentials can be respectively given by, 
\begin{equation}
\Phi_{\text{bulge}} = -{{GM_{\rm b}}\over {\mathcal{R}+q}},
\label{eq:BulgePotential}
\end{equation}
and
\begin{equation}
\Phi_{\text{disk}} = -{{GM_{\rm d}}\over {\sqrt{ \mathcal{R}^2+ (a+\sqrt{z^2+b^2})^2 }}},
\label{eq:DiskPotential}
\end{equation}
where $G$ is the gravitational constant, $M_{\rm b}$ and $M_{\rm d}$ are the mass of the bulge and disk, respectively, and the parameters $q$, $a$, and $b$ correspond to the scale lengths. If $a=0$, the potential of the disk ($\Phi_{\text{disk}}$) simplifies to Plummer's spherical potential \citep[][]{Plummer1911}, while if $b=0$, it simplifies to Kuzmin's potential for a razor-thin disk \citep[][]{Binney2008}. The ratio $b/a$ characterizes the flatness of the M31 disk. Depending on the values of $a$ and $b$, the potential model can range from a nearly infinitesimally thin disk to a spherical system.

For the dark matter halo of M31, we adopt the  NFW model \citep{Navarro1996}, given by, 
\begin{equation}
	\Phi_{\text{halo}} = -{{GM_{\rm vir}\ln{(1+\mathcal{R} c/r_{\rm vir})}}\over{g(c)\mathcal{R}}},
\label{eq:NFWPotential}
\end{equation}
where $r_{\rm vir}$ is the virial radius, $c$ is concentration parameter and $g(c) = \ln(1+c)-{{c}\over {1+c}}$.  $M_{\rm vir}$ is the virial mass given by the equation $M_{\rm vir} = {{4\pi}\over 3}r_{\rm vir}^3\Delta \rho_c$, where $\Delta$ is the virial overdensity parameter, and $\rho_c = {{3H_0^2}\over {8\pi G}}$ is the critical density of the universe, where $H_0 = 70$\,km\,s$^{-1}$\,Mpc$^{-1}$. We set $\Delta$ = 200, which is commonly used in the definition of halo mass in simulations and observations \citep[][]{Seigar2008,Tamm2012,Fardal2013,Veljanoski2014}.

By combining equation~(\ref{eq:BulgePotential}), (\ref{eq:DiskPotential}), and (\ref{eq:NFWPotential}), we can create a model for the total potential of M31. This potential model has seven free parameters. In addition to these parameters, we also have the bulge and halo anisotropy parameters (see Section~\ref{sec:Kinematics Modelling: Bulge and Halo}), which brings the total number of free parameters to nine. We denote these parameters as $\Theta = \{ q,\log(M_{\rm b}), a, b, \log(M_{\rm d}), \log(c), M_{\rm vir},\beta_{\text{bulge}},\beta_{\text{halo}}\}$. We use a Bayesian approach to model the posterior probability distribution of these free parameters, as given by, 
\begin{equation}
	p(\Theta |\mathcal{R}, v_c,v_{c\text{err}}, \sigma_{\text{LOS}}, \sigma_{\text{LOSerr}})  \propto  p(\Theta) \prod \mathcal{L},
\end{equation}
where $v_{c\text{err}}$ is the error of rotation velocity, $\sigma_{\text{LOSerr}}$ is the error of LOS velocity dispersion and $p(\Theta)$ is the prior. We adopt flat priors with the following ranges: $p(q)$, $p(a)$, and $p(b)$ are uniformly distributed between 0 and 10\,kpc, while $p[\log(M_{\rm b}/\text{M}_\odot)]$ and $p[\log(M_{\rm d}/\text{M}_\odot)]$ are uniformly distributed between 9 and 12. We also constrain $\log(M_{\rm b}/\text{M}_\odot)>\log(M_{\rm d}/\text{M}_\odot)$. We consider different previous estimates of the concentration parameters of the Milky Way and M31, as reported in previous studies \citep[e.g.][]{Kafle2014, Huang2016, Kafle2018}, and set $p[\log(c)]$ to be uniformly distributed between $-2$ and 10. Additionally, we set $p[\log(M_{\rm vir}/\text{M}_\odot)]$ to be uniformly distributed between 10 and 14, so the distribution of $M_{\rm vir}/10^{12}\text{M}_\odot$ is $p[x = M_{\rm vir}/10^{12}\text{M}_\odot]= 1/(4 \ln 10 \cdot x)$ between 0.01 and 100.
$\mathcal{L}$ is likelihood, given by, 
\begin{equation}
	\mathcal{L}  = \mathcal{N}(v_c(\mathcal{R})|\text{Model}(\mathcal{R}), v_{c\text{err}}),
\end{equation}
for disk, and
\begin{equation}
	\mathcal{L}  = \mathcal{N}(\sigma_{\text{LOS}}(\mathcal{R})| \text{Model}(\mathcal{R}),\sigma_{\text{LOSerr}}),
\end{equation}
for bulge and halo,
where $\mathcal{N}$ is a Gaussian function, $\Phi(\mathcal{R})$ is the potential.
Model($\mathcal{R}$) is given by equation~\ref{eq:v_candPhi}, while the bulge and the halo $v_c$ is given by equation~\ref{eq:bulgerotation} and equation~\ref{eq:sigmaLOSsigma}. We employ an MCMC analysis to sample prior distribution of these free parameters using the \texttt{emcee} python package \citep{emcee}. 40 walkers are employed and we run the chains for 10,000 steps to ensure convergence, with the first 3,000 steps are discarded as burn-in.

\subsection{The best-fitted potential and mass distribution} 
\label{Sec:potential}

% n: 1.9,1.71,2.67
% Re: 1.93
% gamma: -0.45, -0.12
% alpha: -3.5,-2.2 

% mvir, bulgebeta, halobeta
%1.9,1.93,-0.45,-3.5: 0.89+0.36-0.23, 0.12+0.18-0.37, -0.15+0.12-0.14 %
%1.9,1.93,-0.45,-2.2: 0.86+0.18-0.17, 0.39+0.27-0.41, -0.47+-0.13 %
%1.9,1.93,-0.12,-3.5: 0.84+0.20-0.19, 0.12+0.2-0.33, -0.14+0.12-0.13%
%1.9,1.93,-0.12,-2.2: 0.89+0.33-0.22, 0.08+0.22-0.33, -0.79+0.13-0.11 %
%1.71,1.93,-0.45,-3.5: 0.92+0.17-0.25, 0.48+0.16-0.28, 0.19+0.13-0.13 %
%1.71,1.93,-0.45,-2.2: 0.85+0.24-0.22，0.48+0.17-0.39，-0.45+0.13-0.13 %
%1.71,1.93,-0.12,-3.5: 0.90+0.41-0.25, 0.16+0.2-0.33, -0.12+0.12-0.13 %
%1.71,1.93,-0.12,-2.2: 0.83+0.25-0.20, 0.15+0.17-0.31, -0.77+0.13-0.12 %
%2.67,1.93,-0.45,-3.5: 0.85+0.27-0.22,0.18+0.21-0.38,-0.67+0.12-0.14 %
%2.67,1.93,-0.45,-2.2: 0.88+0.34-0.22，0.43+0.21-0.46，-0.46+0.13-0.14 %
%2.67,1.93,-0.12,-3.5: 0.83+0.19-0.18，0.06+0.2-0.31，-0.15+0.13-0.14 %
%2.67,1.93,-0.12,-2.2: 0.88+0.43-0.22, 0.08+0.21-0.11, -0.78+0.13-0.11 %

Fig.~\ref{fig:PotentialCorner} displays the marginalized (1D) and joint (2D) probability density functions (PDFs) for the model parameters in both one- and two-dimensional (1D and 2D) forms. The joint probability density reveal correlations among some of the parameters, such as a strong correlation between $M_{\rm b}$ and $q$ as well as anticorrelations between $M_{\rm d}$ and $c$. Other pairwise parameters are generally independent. The best-fit values for the model parameters are estimated using the median values of their marginalized PDFs, while uncertainties are calculated from the 68\% probability intervals of the marginalized PDFs. It should be noted that the rotation curve of the disk constrains the rotation of the bulge and halo, effectively determining the values of $\beta_{\text{bulge}}$ and $\beta_{\text{halo}}$. As a result, $\beta_{\text{bulge}}$ and $\beta_{\text{halo}}$ have independent joint distributions with each other and with $M_{\text{vir}}$. However, this does not mean that we can eliminate the degeneracy between mass and anisotropy, as there are uncertainties in the disk rotation curve. For the same reason, $\beta_{\text{bulge}}$ and $\beta_{\text{halo}}$ in the fitting are still recommended values.

According to our analysis, the bulge of M31 has a anisotropy parameter of $\beta_{\text{bulge}} = 0.07^{+0.13}_{-0.29}$. Other kinematic parameters would influence value of $\beta_{\text{bulge}}$, making it varying between $-$0.12 and 0.30. The narrow down process for the uncertainty of $\beta_{\text{bulge}}$ remains challenging. Nevertheless, our calculation leads us to an approximate value of 0, suggesting that M31 likely possesses an isotropic bulge. Regarding the halo, our conclusion indicates a tangential anisotropy parameter, with $\beta_{\text{halo}} = -0.54 ^{+0.38}_{-1.29}$. The MCMC sampling probability distribution of $\beta_{\text{bulge}}$ displays a prolonged negative tail and a corresponding large lower error, indicating a diminishing slope of rotation velocity associated with decreasing $\beta_{\text{bulge}}$.

In our gravitational potential model, the M31 galaxy is divided into the bulge, the disk, and the halo. However, recent studies have suggested that M31 is a barred spiral galaxy \citep[e.g.][]{Athanassoula2006,Beaton2007,Blana2016,Blana2018}. Objects associated with the bar do not follow perfectly circular orbits, which can be misleading when calculating the rotation curve. This can result in a noticeable peak in the rotation curve of the H~{\sc i} gas at around 1-2\,kpc, as shown in Fig.~\ref{fig:DiskRotatingCurve}. It is worth noting that our analysis did not reveal similar structures. The investigations and studies on the bar of M31 primarily focus on the observations in the infrared bands, whereas this particular study employs stars, emission line objects, and star clusters as tracers, utilizing their optical wavelengths spectra. This distinction in observational methods may be a potential factor contributing to the absence of bar structures. Additionally, the rotation velocities of the bulge are determined based on velocity dispersion measurements, rather than being directly derived from observational data. Nevertheless, it remains unclear how this distinction might impact the resulting mass measurement. Another possible explanation for the absence of such structures could be the role of $\beta_{\text{bulge}}$, which effectively constrains the level of the bulge rotation velocities. This limitation may mitigate the impact of the bar on the mass analysis.

In the best-fitted potential model, the dark matter mass of M31 is estimated to be $M_{\rm vir} =  1.14^{+0.51}_{-0.35}\times 10^{12} M_\odot$ within a radius of $r_{\rm vir} = 220 \pm 25$\,kpc. The associated concentration parameter is $\log c = 0.94^{+0.25}_{-0.35}$. In some extent, virial mass scarcely depend on the bulge and halo kinematic parameters because of degeneracy. Virial mass depends on the shape of the rotation curve more and even small oscillations in it, but it is difficult to quantify this effect. Change in kinematic parameters would introduce about an error of 0.05, which is far lower than error from MCMC samples. The resulting mass distribution of M31 is shown in Figure~\ref{fig:CumulativeMass}. Our results are consistent with researches that employed the rotation curve and escape velocity curve methods \citep[e.g.][]{Chapman2006, Tamm2012, Kafle2018}.

\section{Summary}
\label{sec:Summary}

We have obtained a large sample of M31 sources with LOS velocity measurements, consisting of a total of 13,679 objects selected from LAMOST DR9 and the literature. With the largest sample available to date, we have explored the kinematic properties of M31. We have obtained measurements for the radial and tangential velocity dispersion profiles in the disk of M31. Specifically, at a projected radius of $r$ = 10\,kpc, our measurements indicate $\sigma_{R_0} = 62.7^{+3.4}_{-3.4}$\,km\,s$^{-1}$ and $\sigma_{\theta_0} = 69.6^{+3.3}_{-3.3}$\,km\,s$^{-1}$. We also provide the constraints on the scale lengths of both the radial and tangential velocity dispersion, with $L_{R} = 19.8^{+6.5}_{-4.0}$\,kpc and $L_{\theta} = 21.0^{+5.2}_{-3.5}$\,kpc. Our measurements suggest that the velocity dispersion of M31 halo stars exhibits a mild decrease with projected radius, with a power-law index of $\gamma = -0.09 \pm0.03$. We also reveal evidence of a isotropic bulge, with a recommended value $\beta_{\text{bulge}} =  0.07^{+0.13}_{-0.29}$, and tangential anisotropy in the M31 halo, with a recommended value $\beta_{\text{halo}} = -0.54 ^{+0.38}_{-1.29}$.

Based on these derived kinematic properties, together with parameters from existing literature, we have derived the rotation curve of M31 from kinematic models. The rotation curve which encompasses all of the components of M31, including the bulge, disk, and halo, extends up to a distance of about 125\,kpc. The resulting rotation curve of M31 is generally flat at 220\,km\,s$^{-1}$ in the disk, gradually decreasing to 170\,km\,s$^{-1}$ at $\sim$125\,kpc. We have also provided a detailed discussion of the sources of error in our rotation curve measurements, clearly identifying each source of error. Based on the newly derived rotation curve of M31, we have obtained the best estimate of its virial mass to be $M_{\rm vir} = 1.14^{+0.51}_{-0.35} \times 10^{12} M_\odot$ within $r_{\rm vir} = 220 \pm 25$\,kpc. Our measurement is consistent with the results from earlier rotation curve and escape velocity curve analyses. 

However, our understanding of many of the kinematic parameters of M31 remains limited. As large scale multi-object spectroscopic surveys like LAMOST, SDSS, and DESI continue, and new projects like China Space Station Telescope (CSST),  WEAVE (William Herschel Telescope Enhanced Area Velocity Explorer Instrument) %weage
% 4-metre Multi-Object Spectroscopic Telescope (4MOST) 
, and Prime Focus Spectrograph (PFS) in Subaru to be initiated, we anticipate that we will gain a deeper understanding of our neighboring galaxy, much in the same way that we have with the Milky Way.

% pdflatex paper.tex 
% bibtex *.aux 

% e.g. equation~(\ref{eq:quadratic}).
% e.g. Fig.~\ref{fig:example_figure}
% e.g. Table~\ref{tab:example_table}.
% e.g. Section~\ref{sec:maths}

\section*{Acknowledgements}
We express our gratitude for reviewer's thoughtful suggestions. We thank Prof. Yang Huang, Dr. Haifeng Wang and Ms. Yuan Zhou for useful discussions.  This work is partially supported by the National Key R\&D Program of China No. 2019YFA0405500, National Natural Science Foundation of China 11803029, 12173034 and 11833006, Yunnan University Innovation and Entrepreneurship Training Program 202110673079 and Yunnan Province grant No.~C619300A034. We acknowledge the science research grants from the China Manned Space Project with NO.\,CMS-CSST-2021-A09, CMS-CSST-2021-A08 and CMS-CSST-2021-B03. 

This research made use of \texttt{Astropy} \citep{astropy:2013,astropy:2018,astropy:2022}, \texttt{Numpy} \citep{numpy}, \texttt{pandas} \citep{Pandas}, \texttt{Matplotlib} \citep{matplotlib}, \texttt{emcee} \citep{emcee}, and \texttt{corner} \citep{corner}.

Guoshoujing Telescope (the Large Sky Area Multi-Object Fiber Spectroscopic Telescope LAMOST) is a National Major Scientific Project built by the Chinese Academy of Sciences. Funding for the project has been provided by the National Development and Reform Commission. LAMOST is operated and managed by the National Astronomical Observatories, Chinese Academy of Sciences.

\section*{Data availability}

The data underlying this article is available in the manuscript.


\bibliographystyle{mnras}
\bibliography{myreference}

%\begin{thebibliography}{99}
%\end{thebibliography}

\appendix
\section{Derivation of the deprojected circular velocity of halo}
\label{sec:DeprojectedCircularVelocity}

In this section, we present a detailed derivation of the deprojection rotation velocity (equation~\ref{eq:sigmaLOSsigma}). Assuming $\sigma_\theta = \sigma_\phi$, we can express the projected dynamical pressure, which is a product of the observable quantities: the surface density profile $\Sigma (r_p)$ and the LOS velocity dispersion $\sigma_{\text{LOS}}^2$, as follows:

\begin{align}
	\Sigma (r_p)\sigma_{\text{LOS}}^2 & = 2\int_{r_p}^{\infty} [(r^2-r_p^2)\sigma_r^2+r_p^2\sigma_\theta^2]\rho {{dr}\over {r \sqrt{r^2-r_p^2}}} \\
	&= 2\int_{r_p}^{\infty} (1-\beta_{\text{halo}}{{r_p^2}\over {r^2}}) [\rho(r) \sigma_r^2] {{rdr}\over {\sqrt{r^2-r_p^2}}}.
\end{align} 
Here, $\rho(r) \sigma_r^2$ represents the radial dynamical pressure. Thus, $\sigma_{\text{LOS}}$ is given by \citep{Binney1982,mamon2010}:
\begin{equation}
	\sigma_{\text{LOS}}^2(r_p) = {2 \over {\Sigma (r_p)}} \int_{r_p}^{\infty} (1-\beta_{\text{halo}}{{r_p^2}\over {r^2}}) \rho (r) \sigma_r^2{ {rdr}\over{\sqrt{r^2-r_p^2}}}.
\end{equation}
By substituting the halo density profile as shown in Section~\ref{sec:Kinematics Modelling: Bulge and Halo}, we obtain the following equation:
\begin{equation}
	\sigma_{\text{LOS}}^2(r_p) = 2 { {\rho_0}\over {\Sigma_0}}r_p^{-\alpha_{\text{2D}}} \int_{r_p}^{\infty} (1-\beta_{\text{halo}}{{r_p^2}\over {r^2}}) r^{\alpha_{\text{3D}}+1}\sigma_r^2 {{dr}\over{\sqrt{r^2-r_p^2}}}.
	\label{eq:sigmaLOSsigmar2}
\end{equation}
Then we take equation~(\ref{eq:halorotation}) to recover equation ~(\ref{eq:sigmaLOSsigma}) by:
\begin{equation}
	\begin{aligned}
	\sigma_{\text{LOS}}^2(r_p) = &{{2 \times \rho_0 / \Sigma_0}\over {-2\gamma-\alpha_{\text{3D}}-2\beta_{\text{halo}}}} r_p^{-\alpha_{\text{2D}}} \times \\
	&\int_{r_p}^{\infty} (1-\beta_{\text{halo}}{{r_p^2}\over {r^2}}) r^{\alpha_{\text{3D}}+1}v_{c\text{halo}}^2 {{dr}\over{\sqrt{r^2-r_p^2}}}.
	\end{aligned}
\end{equation}
In this equation, only $\frac{\rho_0}{\Sigma_0}$ and $\beta_{\text{halo}}$ are unknown, with the latter being a free parameter in MCMC sampling. We can now calculate $\frac{\rho_0}{\Sigma_0}$. The relationship between surface density $\Sigma (r_p)$ and number density $\rho(r)$ is given by:
\begin{equation}
	\Sigma(r_p) = 2 \int_{r_p}^{\infty} {{\rho(r)rdr}\over{\sqrt{r^2-r_p^2}}}.
\end{equation}
This relationship can be inverted using the Abel transform:
\begin{equation}
	\rho(r) = -{1\over\pi} \int_{r}^{\infty}{{d\Sigma}\over{dr_p}} {{d r_p}\over{\sqrt{r_p^2-r^2}}}.
\end{equation}
By substituting the halo density profile shown in Section~\ref{sec:Kinematics Modelling: Bulge and Halo}, $\frac{\rho_0}{\Sigma_0}$ is calculated as:
\begin{equation}
	{ {\rho_0}\over {\Sigma_0}} = -{{\alpha_{\text{2D}}}\over {\pi}}\int_{1 \text{\,kpc}}^{\infty} r_p^{\alpha_{\text{2D}}-1} {{d r_p}\over{\sqrt{r_p^2-r^2}}} = 0.57 \text{\,kpc}^{-1}.
\end{equation}

\bsp	% typesetting comment
\label{lastpage}
\end{document}